\pdfoutput=1
\documentclass[conference]{IEEEtran}
\ifCLASSINFOpdf
  \usepackage[pdftex]{graphicx}
\else
  \usepackage[dvips]{graphicx}
\fi
\begin{document}
%
% paper title
% can use linebreaks \\ within to get better formatting as desired
% Do not put math or special symbols in the title.
\title{Analysis of Average Travel Time for Stateless Opportunistic Routing Techniques}

% author names and affiliations
% use a multiple column layout for up to three different
%
\author{\IEEEauthorblockN{Sateeshkrishna Dhuli,}
\IEEEauthorblockA{Department of Electrical Engineering,\\
Indian Institute of Technology,\\
Kanpur,\\
Email: dvsk@iitk.ac.in.}
\and
\IEEEauthorblockN{Yatindra Nath Singh,}
\IEEEauthorblockA{Department of Electrical Engineering,\\
Indian Institute of Technology,\\
Kanpur,\\
Email:ynsingh@iitk.ac.in.}
%\thanks{M. Shell is with the Department
%of Electrical and Computer Engineering, Georgia Institute of Technology, Atlanta,
%GA, 30332 USA e-mail: (see http://www.michaelshell.org/contact.html).}% <-this % stops a space
%\thanks{J. Doe and J. Doe are with Anonymous University.}% <-this % stops a space
}

% conference papers do not typically use \thanks and this command
% is locked out in conference mode. If really needed, such as for
% the acknowledgment of grants, issue a \IEEEoverridecommandlockouts
% after \documentclass

% for over three affiliations, or if they all won't fit within the width
% of the page, use this alternative format:
% 
%\author{\IEEEauthorblockN{Michael Shell\IEEEauthorrefmark{1},
%Homer Simpson\IEEEauthorrefmark{2},
%James Kirk\IEEEauthorrefmark{3}, 
%Montgomery Scott\IEEEauthorrefmark{3} and
%Eldon Tyrell\IEEEauthorrefmark{4}}
%\IEEEauthorblockA{\IEEEauthorrefmark{1}School of Electrical and Computer Engineering\\
%Georgia Institute of Technology,
%Atlanta, Georgia 30332--0250\\ Email: see http://www.michaelshell.org/contact.html}
%\IEEEauthorblockA{\IEEEauthorrefmark{2}Twentieth Century Fox, Springfield, USA\\
%Email: homer@thesimpsons.com}
%\IEEEauthorblockA{\IEEEauthorrefmark{3}Starfleet Academy, San Francisco, California 96678-2391\\
%Telephone: (800) 555--1212, Fax: (888) 555--1212}
%\IEEEauthorblockA{\IEEEauthorrefmark{4}Tyrell Inc., 123 Replicant Street, Los Angeles, California 90210--4321}}

% use for special paper notices
%\IEEEspecialpapernotice{(Invited Paper)}

% make the title area
\maketitle

% As a general rule, do not put math, special symbols or citations
% in the abstract
\begin{abstract}
Wireless network applications, such as, searching, routing, self stabilization and query processing can be modeled as random walks on graphs. Stateless Opportunistic routing technique is a robust distributed routing technique based on random walk approach , where nodes transfer the packets to one of their 
direct neighbors uniformly, until the packets reach their destinations. Simplicity in execution, fault tolerance, low overhead and robustness to topology changes made it more suitable to wireless sensor networks scenarios.  But the main disadvantage of stateless opportunistic routing is estimating and studying the effect of network parameters on the packet latency. In this work, we derived the analytical expressions for mean latency or average packet travel time for $r$-nearest neighbor cycle, $r$-nearest neighbor torus networks. Further, we derived the generalized expression for mean latency for $m$-dimensional $r$-nearest neighbor torus networks and studied the effect of number of nodes, nearest neighbors and network dimension on average packet travel time.
\end{abstract}

% Note that keywords are not normally used for peerreview papers.
\begin{IEEEkeywords}
Wireless sensor networks, Delay tolerant networks, Random walks, Opportunistic forwarding, Spectral graph theory
\end{IEEEkeywords}

% no keywords

% For peer review papers, you can put extra information on the cover
% page as needed:
% \ifCLASSOPTIONpeerreview
% \begin{center} \bfseries EDICS Category: 3-BBND \end{center}
% \fi
%
% For peerreview papers, this IEEEtran command inserts a page break and
% creates the second title. It will be ignored for other modes.
\IEEEpeerreviewmaketitle

\section{Introduction}

\IEEEPARstart{I}{n} Stateless opportunistic routing, packets are forwarded to the next available neighbors in a random walk fashion until they reach the destinations. Estimation of access time, commute time, cover time and mixing time for random walks are discussed in \cite{randsur}. Opportunistic forwarding provides significant performance gains and increases the throughput in the wireless networks \cite{routing}, \cite{Oppor}. The problem of searching for a node or a piece of data and the hitting time of the node has been studied in \cite{time}. Analytic formulas for maximum expected latency has been derived for regular wireless networks \cite{analysis}. But this work does not studied the Mean Latency metric which is very important metric to study the packet delay and it also does not provide the upper and lower bounds for latency. In our work we derived the mean latency expressions for $r$-nearest neighbor networks. The motivation behind the using finite sized networks is most of the practical  WSN/adhoc networks are finite sized, such as applications in health, military and security in buildings. The $r$-nearest neighbor networks \cite{analysis} with varying number of nodes represents the notion of geographical proximity in the wireless sensor networks/ adhoc networks, where, nearest neighbors $r$ captures the overhead or nodes' transmission radius. The advantage of this kind of analysis and theoretical results is they will play a critical role in the design of wireless sensor networks before the network operations, and also easier to perform than real experiments and thousands of simulation trails. This work provides the understanding of mean latency in terms of number of nodes, nearest neighbors and network dimension and gives the important insights for estimating latency in wireless networks. Further, we also studied the effect of wireless network parameters on packet delay in flat fading environments. For that, we used the system model proposed in \cite{Optimal} for designing the topology coefficients.

The rest of the paper is organized as follows. In Section II, we have given a brief overview about Mean Latency or average travel time. In Section III, we have given the generalized expressions for eigen values of the Laplacian matrix for $r$-nearest neighbor networks. In Section IV, Mean Latency expressions and bounds are derived for 
$r$-nearest neighbor cycle, $r$-nearest neighbor torus and $m$-dimensional $r$-nearest neighbor networks have been derived. In Section V, we have studied the network parameters effect on latency for arbitrary network model. In Section VI, we compared the simulation results with analytical results obtained in the Section IV.

\section{Mean Latency of Random Walks}
Given an undirected graph $G = (V,E)$, where $V$ is the set of nodes and $E$ is the set of edges. Let $A$ be a adjacency matrix of $G$ and $d_i$ be the degree of $i$ where each node $i \in V$. Let $D = diag\left( {d_i } \right)$ be a diagonal matrix of node degrees, then $P = D^{ - 1} A $ is a symmetric transition matrix associated with a random walk on $G$. Let $\pi  = \left[ {\pi _i } \right]_{1 \le i \le n}$ is a stationary distribution probability vector. In this case random walk on $G$ is reversible,.i.e. $\pi _i p_{ij}  = \pi _j p_{ji}$ and distribution can be expressed as

\begin{equation}
\pi _i  = \frac{{d_i }}{{\sum\nolimits_k {d_k } }} = \frac{{d_i }}{d}
\label{1}
\end{equation}

Definition 1: Normalized Laplacian matrix for undirected graph $G$ is defined as

\begin{equation}
N = D^{ - \frac{1}{2}} (D - A)D^{ - \frac{1}{2}}
\label{2}
\end{equation}

Where $N$ is symmetric and positive semi-definite. Let $\lambda _k$, $v_k$ be the eigen values and the corresponding eigen vectors of $N$, then  the hitting time of random walk \cite{randsur} from node $s$ to node $t$ as $H_{st}$, which can be expressed as  

\begin{equation}
H_{st} = 2m\sum\limits_{k = 2}^n {\frac{1}{{\lambda _k }}} \left( {\frac{{v_{kt}^2 }}{{d(t)}} - \frac{{v_{ks} v_{kt} }}{{\sqrt {d(t)d(s)} }}} \right)
\label{3} 
\end{equation}

$Lemma$ 1: The mean latency or average random walk travel time $T$ \cite{robust} between every arbitrary pair of nodes is equal to 
\begin{equation}
T=\frac{2}{{n - 1}}Tr(L^ +)
\label{4}
\end{equation}
Where $Tr(L^ +)$ represents trace of the Moore-Penrose inverse of Laplacian matrix and $n$ denotes the number of nodes.

\section{$r$-nearest neighbor networks}

\subsection{$r$-nearest neighbor cycle}
The $r$-nearest neighbor cycle $C_n^r$ can be represented by a circulant matrix \cite{Circul}. A circulant matrix is defined as
 
\begin{equation}
\left[ \begin{array}{l}
 a_1 \,\,a_2 \,\,........a_{n - 1} \,\,a_n  \\ 
 a_n \,\,a_1 \,\,....\,....a_{n - 2} \,a_{n - 1}  \\ 
 .\,\,\,\,\,\,\,.\,\,\,\,\,\,\,\,\,\,\,\,\,\,\,\,\,\,\,\,\,\,\,\,.\,\,\,\,\,\,\,\,\,. \\ 
 .\,\,\,\,\,\,\,.\,\,\,\,\,\,\,\,\,\,\,\,\,\,\,\,\,\,\,\,\,\,\,\,.\,\,\,\,\,\,\,\,\,. \\ 
 a_3 \,\,a_4 \,\,\,...........a_1 \,\,\,a_2  \\ 
 a_2 \,\,a_3 \,\,.............a_n \,\,a_1 \,\, \\ 
 \end{array} \right]
 \label{5}
\end{equation}

and $j$-th eigen value  of a circulant matrix can be expressed as

\begin{equation}
\lambda _j  = a_1  + a_2 \omega ^{j}  + .............. + a_n \omega ^{(n - 1)j} 
\label{6}
\end{equation}

where $\omega$ be the $n$-th root of 1. Then $\omega$ is the complex number: 
\begin{equation}
\omega  = \cos \left( {\frac{{2\pi }}{n}} \right) + i\sin \left( {\frac{{2\pi }}{n}} \right)= e^\frac{i2\pi}{n}
\label{7}
\end{equation}

The 1-nearest cycle and 2-nearest cycle are shown in Fig.1 and Fig.2 respectively. Let the adjacency matrix $A$ and the degree matrix $D$ of 1-nearest cycle, then they can be written as

\begin{equation}
A=\left[ \begin{array}{l}
 0\,\,1\,\,0\,\,..............0\,\,1 \\ 
 1\,\,0\,\,1\,\,..............0\,0 \\ 
 .\,\,\,\,.\,\,\,\,.\,\,\,\,\,\,\,\,\,\,\,\,\,\,\,\,\,\,\,\,\,\,\,\,\,.\,\,\,. \\ 
 .\,\,\,\,.\,\,\,\,.\,\,\,\,\,\,\,\,\,\,\,\,\,\,\,\,\,\,\,\,\,\,\,\,\,.\,\,\,. \\ 
 0\,\,0\,\,0\,\,..............0\,\,1 \\ 
 1\,\,0\,\,0\,\,..............1\,0\, \\ 
 \end{array} \right]
 \label{8}
\end{equation}

\begin{equation}
D=\left[ \begin{array}{l}
 2\,\,0\,0\,\,............0\,\,0 \\ 
 0\,\,2\,0\,\,....\,........0\,\,0 \\ 
 .\,\,\,\,.\,\,\,\,\,\,\,\,\,\,\,\,\,\,\,\,\,\,\,\,\,\,\,\,\,\,\,\,\,.\,\,\,. \\ 
 .\,\,\,\,.\,\,\,\,\,\,\,\,\,\,\,\,\,\,\,\,\,\,\,\,\,\,\,\,\,\,\,\,\,.\,\,\,. \\ 
 0\,\,0\,\,\,...............2\,0\,\,\, \\ 
 0\,\,0\,\,\,...............0\,\,2 \\ 
 \end{array} \right]
 \label{9}
\end{equation}

$Theorem$ 1: The generalized expression for eigenvalues of Laplacian matrix $L$ for $r$-nearest neighbor cycle $C_n^r $ can be expressed as,
\begin{equation}
\lambda _j (L(C_n^r))\, = 2r + 1 - \frac{{\sin \frac{{(2r + 1)\pi j}}{n}}}{{\sin \frac{{\pi j}}{n}}}
\label{10}
\end{equation}
where $j=0,1,...(n-1)$.\\
$Proof$:  From (\ref{7}), we can observe that, the first row is enough to obtain the eigen values of any circulant matrix.

The first row of adjacency matrix (A), degree matrix (D) and Laplacian matrix (L) can be written as follows,
\begin{equation}
A_{1n}  = \left[ {0\underbrace {\,1\,\,1\,\,1\,........0..........\,1\,\,1\,\,1}_{2r\,times}\,} \right]
\label{11}
\end{equation} 

\begin{equation}
D_{1n}  = \left[ {2r\,\,0\,\,0\,\,0\,.....0\,\,0\,\,0\,\,0} \right]
\label{12}
\end{equation}

\begin{equation}
\resizebox{.9 \hsize} {!} {$L_{1n}  = \left[ {2r\,\,\underbrace { - 1\,\, - 1\,\, - 1\,.....0...\,.... - 1\, - 1\, - 1}_{2r\,times}} \right]$}
\label{13}
\end{equation}

By using (\ref{13}) and (\ref{6}), we can write the   

\begin{equation} 
\lambda _j (L(C_n^r)) = 2r - 2\sum\limits_{i = 1}^r {\cos \left( {\frac{{2\pi ji}}{n}} \right)}
\label{14}
\end{equation}

Lemma 2: Trigonometric identity of Dirichlet kernel \cite{ident}
\begin{equation}
1 + 2\sum\limits_{j = 1}^r {\cos (jx)}  = \frac{{\sin \left( {r + \frac{1}{2}} \right)x}}{{\sin \left( {\frac{x}{2}} \right)}}
\label{15}
\end{equation}

Hence, from the Lemma 2, (\ref{14}) can be rewritten as,

\begin{equation}
\lambda _j (L(C_n^r))\, = 2r + 1 - \frac{{\sin \frac{{(2r + 1)\pi j}}{n}}}{{\sin \frac{{\pi j}}{n}}} \nonumber
\end{equation}

\begin{figure}[!t]
\centering
\includegraphics[width=2.2in]{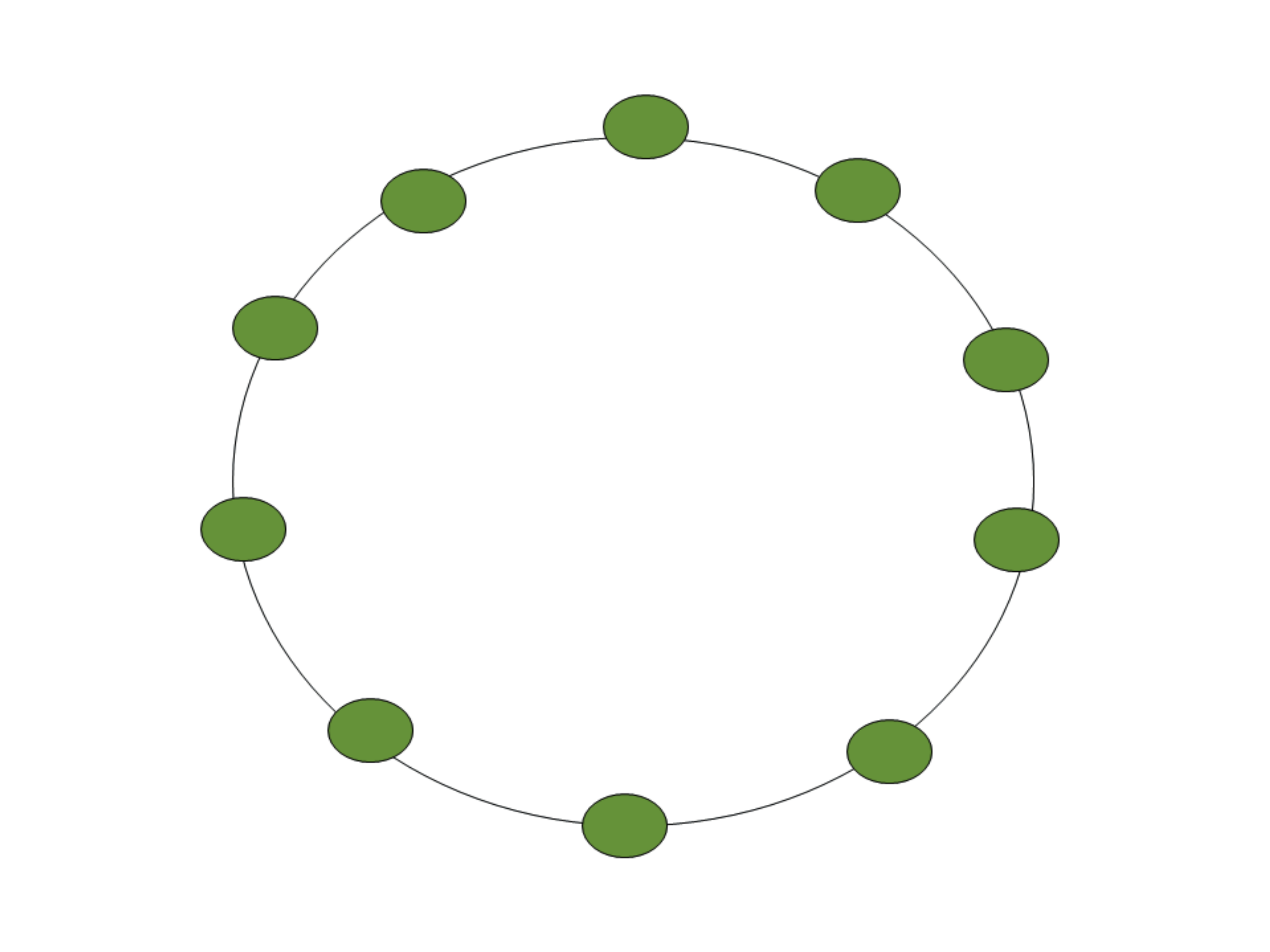}
\caption{1-nearest neighbor cycle }
\label{fig:1}
\end{figure}

\begin{figure}[!t]
\centering
\includegraphics[width=2in]{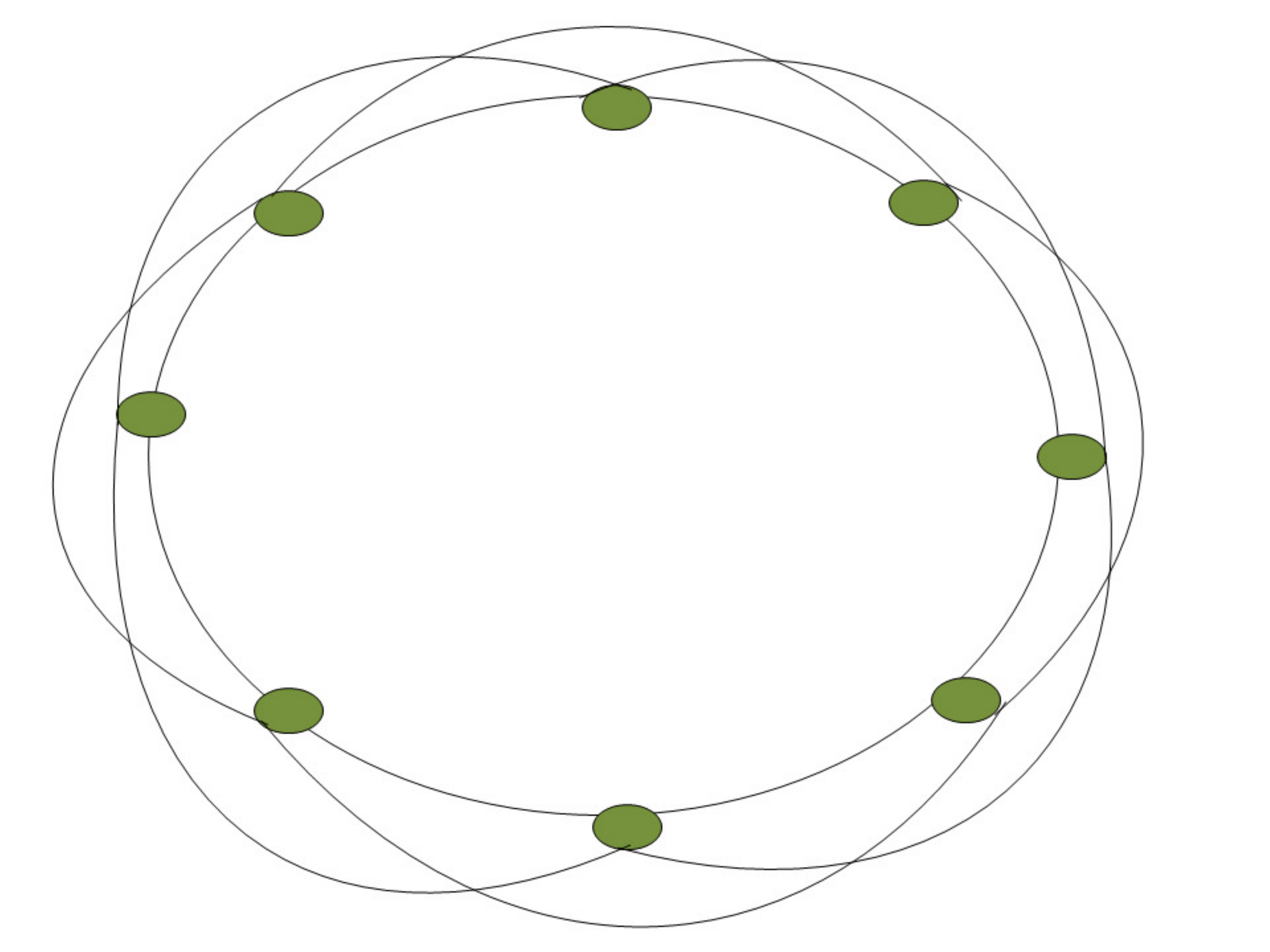}
\caption{2-nearest neighbor cycle}
\label{fig:2}
\end{figure}

\subsection{$r$-nearest neighbor torus}

A torus can be seen in Fig. 3 and it can be represented by the $n \times n$ block circulant matrix $A$ as
\begin{equation}
A = \left[ \begin{array}{l}
 A_0 \,\,\,\,\,\,\,\,\,\,A_1 \,\,........A_{n_1 - 2} \,\,A_{n_1 -1}  \\ 
 A_{n_1 -1} \,\,A_0 \,\,....\,....A_{n_1 - 3} \,A_{n_1 - 2}  \\ 
 .\,\,\,\,\,\,\,\,\,\,\,\,\,\,\,\,\,\,\,\,\,.\,\,\,\,\,\,\,\,\,\,\,\,\,\,\,\,\,\,\,\,\,\,\,\,.\,\,\,\,\,\,\,\,\,\,\,\,\,\,\,. \\ 
 .\,\,\,\,\,\,\,\,\,\,\,\,\,\,\,\,\,\,\,\,\,.\,\,\,\,\,\,\,\,\,\,\,\,\,\,\,\,\,\,\,\,\,\,\,\,.\,\,\,\,\,\,\,\,\,\,\,\,\,\,\,. \\ 
 A_1 \,\,\,\,\,\,\,\,\,\,\,\,\,A_2 \,\,..........A_{n_1 -1} \,\,A_0 \,\, \\ 
 \end{array} \right]
 \label{16}
\end{equation}

where the number of nodes $n=n_1^2$, then each block $A_i$, for $i=0,1...(n_1-1)$ represents $n_1 \times n_1$ circulant matrices.\\

$Lemma$ 2: Let $G$ be the cartesian product of two graphs $G^{'}$ and $G^{''}$ with vertex sets $V^{'}$ and $V^{''}$ and edge sets $E^{'}$ and $E^{''}$.
Let the eigen values of $G^{'}$ are  $\lambda _1 \left( {G^{'} } \right)..........\lambda _p \left( {G^{'}} \right)$ and $G^{''}$ are $\lambda _1 \left( {G^{''} } \right)..........\lambda _q \left( {G^{''} } \right)$, where $ p = \left| {V^{'} } \right|$ and $q = \left| {V^{''} } \right|$. Let the vertex set of $G$ is $r = \left| {V } \right|$, which can be expressed as $V = \left| {V^{'} } \right| \times \left| {V^{''} } \right|$ \cite{spectra}. Then, the eigen values of $G$ can be expressed as 
\begin{equation}
\lambda _k \left( G \right) = \lambda _i \left( {G^{'} } \right) + \lambda _j \left( {G^{''} } \right)
\label{17}
\end{equation}
,where $i \in \{ 1,2,....p\}$, $j \in \{ 1,2,....q\}$ and $k \in \{ 1,2,....r\}$. \\

$Remark$ 2: (\ref{17}) also holds for eigen values of the Laplacians $L^{'}$ and $L^{''}$ of graphs of $G^{'}$ and $G^{''}$ respectively\cite{opt}.\\

$Theorem$ 2: The generalized expression for eigenvalues of Laplacian matrix $L$ for $r$-nearest neighbor torus $T_n^r$ can be expressed as
\begin{equation}
\lambda _{j_1 ,j_2 } \left( {L(T_{k_1 ,k_2 }^r )} \right) = 4r + 2 - \frac{{\sin \frac{{(2r + 1)\pi j_1 }}{{k_1 }}}}{{\sin \frac{{\pi j_1 }}{{k_1 }}}} - \frac{{\sin \frac{{(2r + 1)\pi j_2 }}{{k_2 }}}}{{\sin \frac{{\pi j_2 }}{{k_2 }}}}
\label{18}
\end{equation}

where $j_1  = 0,1,2,...(k_1  - 1), j_2  = 0,1,2,...(k_2  - 1)$.\\

$Proof$ :  
$T_n^r$ can be represented by Cartesian product of two $r$-nearest neighbor cycles.
So from the $Lemma$ 2, we can write the $\lambda _{j_1 ,j_2 } \left( {L(T_{k_1 ,k_2 }^r )} \right)$ as,

\begin{equation}
\lambda _{j_1 ,j_2 } \left( {L(T_{k_1 ,k_2 }^r )} \right) = \lambda _{j_1 } \left( {L(C_{k_1 }^r )} \right) + \lambda _{j_2 } \left( {L(C_{k_2 }^r )} \right)
\label{19}
\end{equation}

From $Remark$ 2, we can write the expressions for $\lambda _{j_1 } \left( {L(C_{k_1 }^r )} \right)$ and  $\lambda _{j_2 } \left( {L(C_{k_2 }^r )} \right)$, substituting them in (\ref{19}) proves the theorem.
\\

$Theorem$ 3: The generalized expression for eigenvalues of Laplacian matrix $L$ for $m$-dimensional $r$-nearest neighbor torus can be expressed as

\begin{equation}
\resizebox{.9 \hsize} {!} {$\lambda _{j_1 ,j_2,... j_m} \left( {L(T_{k_1 ,k_2....k_m}^r )} \right) = (2r + 1)m - \sum\limits_{i = 1}^m {\left( {\frac{{\sin \frac{{(2r + 1)\pi j_i }}{{k_i }}}}{{\sin \frac{{\pi j_i }}{{k_i }}}}} \right)}$}
\label{20}
\end{equation}

$Proof$: 
$r$-nearest neighbor $m$-dimensional torus can be represented by Cartesian product of $m$ number of $r$-nearest neighbor cycles.
So from the $Lemma$ 2, we can write the $\lambda _{j_1 ,j_2....j_m } \left( {L(T_{k_1 ,k_2....k_m }^r )} \right)$ as,

\begin{equation}
\resizebox{.9 \hsize} {!} {$\lambda _{j_1 ,j_2,... j_m} \left( {L(T_{k_1 ,k_2....k_m}^r )} \right) = \lambda _{j_1 } \left( {L(C_{k_1 }^r )} \right) +\lambda _{j_2 } \left( {L(C_{k_1 }^r )} \right) .............+\lambda _{j_m } \left( {L(C_{k_m }^r )} \right)$}
\label{21}
\end{equation}

From $Remark$ 2, we can substitute the expressions for $\lambda _{j_1 } \left( {L(C_{k_1 }^r )} \right)$, $\lambda _{j_2 } \left( {L(C_{k_2 }^r )} \right)$ and $\lambda _{j_m } \left( {L(C_{k_m }^r )} \right)$ in (\ref{21}), which proves the theorem.\\
\\

\begin{figure}[!t]
\centering
\includegraphics[width=2.2 in]{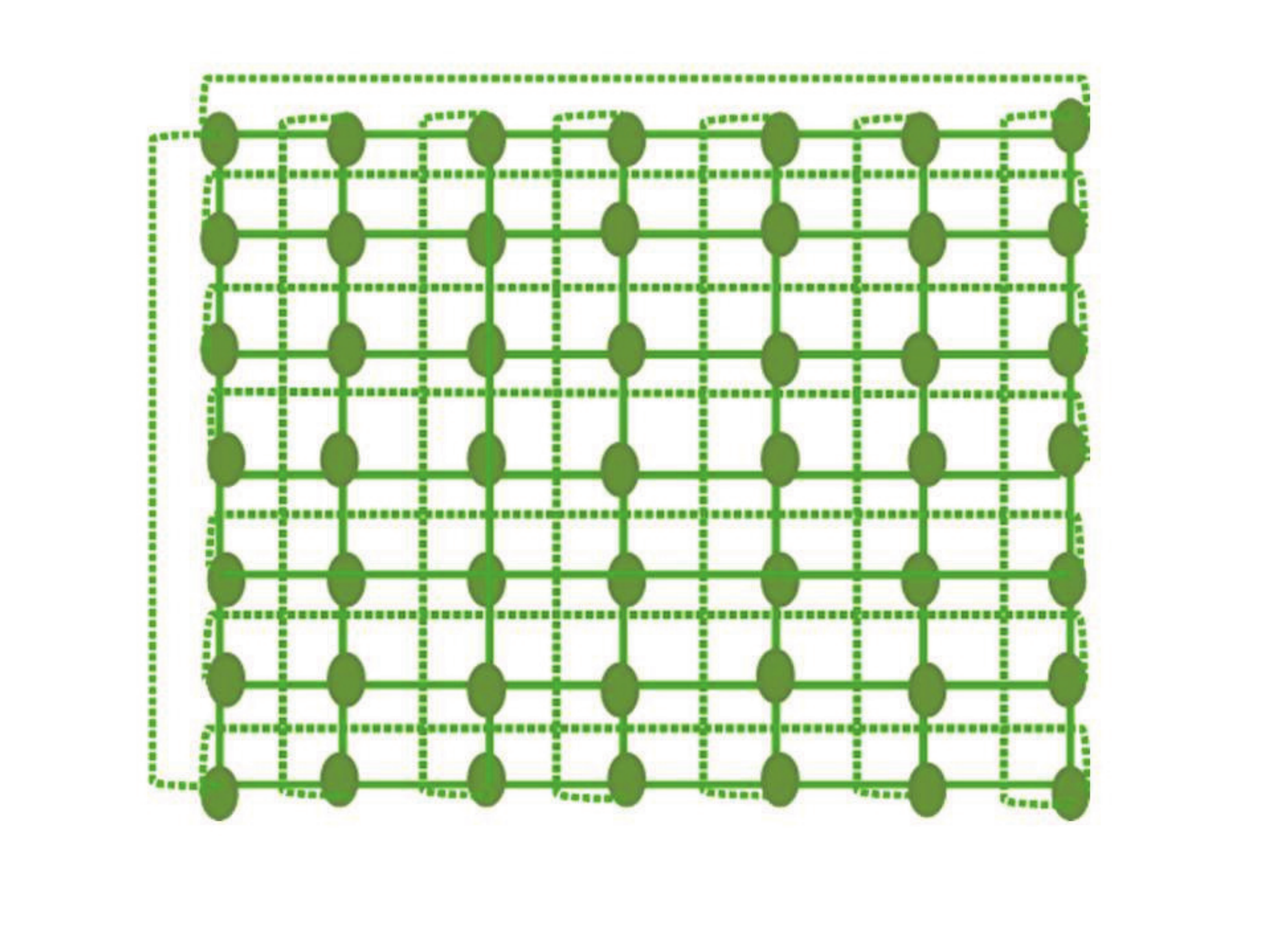}
\caption{Two dimensional torus }
\label{fig:3}
\end{figure}

\section{Mean latency analysis for $r$-nearest neighbor networks}
 
$Theorem$ 5: The mean latency $T$ of $r$-nearest neighbor cycle ${C_n^r }$ between every arbitrary pair of nodes is 

\begin{equation}
\resizebox{.9 \hsize} {!} {$T\left( {C_n^r } \right) = \sum\limits_{j = 2}^n {\left( {\frac{2}{{(n - 1)\left( {(2r + 1) - \frac{{\sin \frac{{(2r + 1)\pi j}}{n}}}{{\sin \frac{{\pi j}}{n}}}} \right)}}} \right)}$}
\label{23}
\end{equation}
Proof:
From (\ref{10}), we can write 
\begin{equation}
\begin{array}{l}
 Tr(L^ +  ) = \sum\limits_{j = 2}^n {\frac{1}{{\lambda _j (L)}}}  \\ 
 \,\,\,\,\,\,\,\,\,\,\,\,\,\,\, = \sum\limits_{j = 2}^n {\left( {\frac{1}{{\left( {(2r + 1) - \frac{{\sin \frac{{(2r + 1)\pi j}}{n}}}{{\sin \frac{{\pi j}}{n}}}} \right)}}} \right)}  \\ 
 \end{array}
 \label{24}
\end{equation}

Substituting the (\ref{24}) in (\ref{4}), proves the theorem.

$Theorem$ 6: The mean latency of $r$-nearest neighbor torus $T_{k_1 ,k_2 }^r$ between every arbitrary pair of nodes is 
\begin{equation}
\resizebox{.9 \hsize} {!} {$T(T_{k_1 ,k_2 }^r ) = \sum\limits_{j_1  = 1}^{k_1  - 1} {\sum\limits_{j_2  = 0}^{k_2  - 1} {\left( {\frac{2}{{\left( {k_1  + k_2  - 1} \right)\left( {(4r + 2) - \frac{{\sin \frac{{(2r + 1)\pi j_1 }}{{k_1 }}}}{{\sin \frac{{\pi j_1 }}{{k_1 }}}} - \frac{{\sin \frac{{(2r + 1)\pi j_2 }}{{k_2 }}}}{{\sin \frac{{\pi j_2 }}{{k_2 }}}}} \right)}}} \right)}}$}
\label{25}
\end{equation}
Proof:
From (\ref{18}), we can write 
\begin{equation}
\begin{array}{l}
 Tr(L^ +  ) = \sum\limits_{j_1  = 1}^{k_1  - 1} {\sum\limits_{j_2  = 0}^{k_2  - 1} {\frac{1}{{\lambda _{j_1 ,j_2 } (L)}}} }  \\ 
 \,\,\,\,\,\,\,\,\,\,\,\,\,\,\, = \sum\limits_{j_1  = 1}^{k_1  - 1} {\sum\limits_{j_2  = 0}^{k_2  - 1} {\left( {\frac{2}{{\left( {(4r + 2) - \frac{{\sin \frac{{(2r + 1)\pi j_1 }}{{k_1 }}}}{{\sin \frac{{\pi j_1 }}{{k_1 }}}} - \frac{{\sin \frac{{(2r + 1)\pi j_2 }}{{k_2 }}}}{{\sin \frac{{\pi j_2 }}{{k_2 }}}}} \right)}}} \right)} }  \\ 
 \end{array}
 \label{26}
\end{equation}
Substituting the (\ref{26}) in (\ref{4}), proves the theorem.

$Theorem$ 7: The mean latency of $r$-nearest neighbor torus $T_{k_1 ,k_2 ,....k_m }^r $ between every arbitrary pair of nodes is 
\begin{equation}
\resizebox{.9 \hsize} {!} {$T(T_{k_1 ,k_2 ,....k_m }^r ) = \sum\limits_{j_1  = 1}^{k_1  - 1} {\sum\limits_{j_2  = 0}^{k_2  - 1} {......\sum\limits_{j_m  = 0}^{k_m  - 1} {\left( {\frac{2}{{\left( {(2r + 1)m - \sum\limits_{i = 1}^m {\frac{{\sin \frac{{(2r + 1)\pi j_i }}{{k_i }}}}{{\sin \frac{{\pi j_i }}{{k_i }}}}} } \right)}}} \right)} } }$}
\label{27}
\end{equation}
Proof:
From (\ref{20}), we can write 
\begin{equation}
\resizebox{.9 \hsize} {!} {$\begin{array}{l}
 Tr(L^ +  ) = \sum\limits_{j_1  = 1}^{k_1  - 1} {\sum\limits_{j_2  = 0}^{k_2  - 1} {......\sum\limits_{j_m  = 0}^{k_m  - 1} {\frac{1}{{\lambda _{j_1 ,j_2 .....j_m } (L)}}} } }  \\ 
 \,\,\,\,\,\,\,\,\,\,\,\,\,\,\, = \sum\limits_{j_1  = 1}^{k_1  - 1} {\sum\limits_{j_2  = 0}^{k_2  - 1} {......\sum\limits_{j_m  = 0}^{k_m  - 1} {\left( {\frac{2}{{\left( {\sum\limits_{i = 1}^m {k_i  - 1} } \right)\left( {(2r + 1)m - \sum\limits_{i = 1}^m {\frac{{\sin \frac{{(2r + 1)\pi j_i }}{{k_i }}}}{{\sin \frac{{\pi j_i }}{{k_i }}}}} } \right)}}} \right)} } }  \\ 
 \end{array}
 $}
 \label{28}
\end{equation}
Substituting the (\ref{28}) in (\ref{4}), proves the theorem.

$Lemma$ 3: The mean latency $T$ between every arbitrary pair of nodes satisfies the following bound \cite{survival}:

\begin{equation}
\frac{2}{{(n - 1)\lambda _1 (L)}} \le T \le \frac{2}{{\lambda _1 (L)}}
\label{29}
\end{equation}

Where $\lambda _1 (L)$ is the second smallest eigen value of Laplacian matrix and $n$ represents number of nodes.

$Theorem$ 8: The bounds for mean latency $T$ for $r$-nearest neighbor cycle can be expressed as,

\begin{equation}
\resizebox{.9 \hsize} {!} {$\frac{{2\sin \frac{\pi }{n}}}{{(n - 1)((2r + 1)\sin \frac{\pi }{n} - \sin \frac{{(2r + 1)\pi }}{n})}} \le T \le \frac{{2\sin \frac{\pi }{n}}}{{(2r + 1)\sin \frac{\pi }{n} - \sin \frac{{(2r + 1)\pi }}{n}}}$}
\label{30}
\end{equation}
Proof:
Substituting the n=1 in (\ref{4}) gives $\lambda _1 (L)$, which can be substituted in the (\ref{29}) proves the theorem.

Similarly we can prove the bounds for mean latency for $m$-dimensional torus network as

\begin{equation}
\resizebox{.9 \hsize} {!} {$\frac{{2\sin \frac{\pi }{{k_1 }}}}{{(n - 1)((2r + 1)\sin \frac{\pi }{{k_1 }} - \sin \frac{{(2r + 1)\pi }}{{k_1 }})}} \le T \le \frac{{2\sin \frac{\pi }{{k_1 }}}}{{(2r + 1)\sin \frac{\pi }{{k_1 }} - \sin \frac{{(2r + 1)\pi }}{{k_1 }}}}$}
\label{31}
\end{equation}

\section{Opportunistic Forwarding for arbitrary networks}
We consider the arbitrary network model as shown in Fig. 9, where the nodes are distributed arbitrarily, but their positions are known. The following propagation model has been considered 

\begin{equation}
P_{R_j }  = \frac{{p_{ij} }}{{1 + \left( {{{r_{ij} } \mathord{\left/
 {\vphantom {{r_{ij} } {r_0 }}} \right.
 \kern-\nulldelimiterspace} {r_0 }}} \right)^\eta  }}
 \label{32}
\end{equation}

where $P_{R_j }$ is the power received by node $j$ when node $i$ transmits, $r_{ij}$ is the distance between nodes $i$ and $j$, $\eta$ is the path loss exponent and $r_{0}$ is a reference distance. If $r_{ij}  \gg r_0$, the receiver is in the transmit antenna far-field, where the received power is inversely proportional to $r_{ij}^\eta$. Conversely, if  $r_{ij}  \ll r_0$, then the received power is approximately equal to the transmitted power.

The relationship between the coefficients $a_{ij}$ and the distances $r_{ij}$ can be expressed as

\begin{equation}
a_{ij}  = \frac{1}{{1 + \left( {{{r_{ij} } \mathord{\left/
 {\vphantom {{r_{ij} } {r_{c_{ij} } }}} \right.
 \kern-\nulldelimiterspace} {r_{c_{ij} } }}} \right)^{^\alpha  } }}
 \label{33}
\end{equation}

where $\alpha$ is a positive coefficient and $r_{c_{ij}}$ is the coverage radius, which depends on the transmit power. 

%According to (\ref{eq:3}) we can get the following relations
%
%\begin{equation}\label{eq:4}
%a_{ij}  = \left\{ \begin{array}{l}
% 1\,\,\,\,\,if\,\,r_{ij}  \ll {r_{c_{ij} }}  \\ 
% 0\,\,\,if\,\,\,r_{ij}  \gg {r_{c_{ij} }} \, \\ 
% \end{array} \right.
%\end{equation}

From the (\ref{33}), the relationship between the coverage radius $r_{c_{ij}}$, power coefficients $p_{ij}$ and the minimum required power for communication $p_{min}$ can be expressed as

\begin{equation}
r_{c_{ij} }  = r_0 \left( {\frac{{p_{ij} }}{{p_{\min } }} - 1} \right)^{{\raise0.7ex\hbox{$1$} \!\mathord{\left/
 {\vphantom {1 \eta }}\right.\kern-\nulldelimiterspace}
\!\lower0.7ex\hbox{$\eta $}}} 
\label{34}
\end{equation}

where 

\begin{equation}
r_0=\sqrt{\frac{log(n)+c_n}{\pi n}}
\label{35}
\end{equation}

using the (\ref{34}) and (\ref{32}) the topology coefficients can be written in terms of the power coefficients $p_{ij}$, denoting the power used by node $i$ to transmit to node $j$ as

\begin{equation}
a_{ij}  = \frac{{r_0^\alpha  \left( {p_{ij}  - p_{\min } } \right)^{{\raise0.7ex\hbox{$\alpha $} \!\mathord{\left/
 {\vphantom {\alpha  \eta }}\right.\kern-\nulldelimiterspace}
\!\lower0.7ex\hbox{$\eta $}}} }}{{r_0^\alpha  \left( {p_{ij}  - p_{\min } } \right)^{{\raise0.7ex\hbox{$\alpha $} \!\mathord{\left/
 {\vphantom {\alpha  \eta }}\right.\kern-\nulldelimiterspace}
\!\lower0.7ex\hbox{$\eta $}}}  + r_{ij}^\alpha  p_{\min }^{{\raise0.7ex\hbox{$\alpha $} \!\mathord{\left/
 {\vphantom {\alpha  \eta }}\right.\kern-\nulldelimiterspace}
\!\lower0.7ex\hbox{$\eta $}}} }}
\label{36}
\end{equation}

Remark 1 : From the (\ref{36}), it is evident that the topology coefficients depends on power coefficients $p_{ij}$ and other wireless network parameters $\alpha$, $p_{min}$, $r_{0}$, $r_{ij}$,  $\eta$. 

Remark 2:  From the (\ref{36}) it is also evident that symmetric wireless links present in the network when $p_{ij} = p_{ji}$ and asymmetric wireless links exits when $p_{ij} \neq p_{ji}$. \\
Here, we studied the symmetrical wireless networks.

\section{Simulation results}
To validate the derived Mean latency analytical expressions, we compared with the simulation results and observed that both agree with each other. As shown in the Fig. 4, we plotted mean latency $T$ against nearest neighbors for $n$=300, and observed that mean latency decreases with $r$. From Fig.5, we can see the mean latency versus number of nodes $n$ for $r=1$ for $r$-nearest neighbor cycle. Similarly to observe the Mean latency variation for two dimensional finite networks, we plotted mean latency versus $k_{1}$ and $k_{2}$. We have taken $r$=1 to plot Fig. 7. and $k_{1}$=$k_{2}$=1000 to plot Fig.6. To study the effect of network dimension on mean latency, we have taken the $k_{1}$=16, $k_{2}$=18, $k_{3}$=20 and $k_{4}$=22 and $r$ varied from $1$ to $4$. We have observed that, Mean latency is decreased with the network dimension and it further decreases with increase in $r$.

To study the effect of few more wireless network parameters on expected packet delay (EPD), we have used the (\ref{3}) and to generate the topologies as shown in Fig.9 we have used the (\ref{36}). 

\begin{equation}
EPD =\frac{\sum_{i=1}^{n}\sum_{j=1}^{n} H_{ij}}{n(n-1)}
\label{36}
\end{equation}
where $H_{ij}$ denotes the delay between node $i$ and node $j$ and $n$ represents number of nodes.

After estimating the EPD, we try to understand the effect of wireless network parameters on EPD.To evaluate how the network density affects the EPD, topologies has been generated with size, $A = 1 \times 1$ $ m^2$. From Fig. 10, we can see that, packet delay increases with path loss exponent, here we have taken $p_{min}=0.1$ and $n=30$. Fig. 11 shows the impact of Minimum Received Power on EPD for $n=30$ in freespace ($\alpha=2$) and multipath ($\alpha=4$) communication. From the simulation results, EPD increases with Minimum  Received Power ($p_{min}$) and it increases further in multipath environments.  From the proposed analytic  modeling, topological coefficients has been derived and the connectivity threshold has been introduced to get the binary adjacency matrix which defines the wireless network topology. Fig. 12 shows the impact of connectivity threshold on EPD for $p_{min}=0.1$ in freespace ($\alpha=2$) and multipath ($\alpha=4$) communication.

\begin{figure}[!t]
\centering
\includegraphics[width=2.5 in]{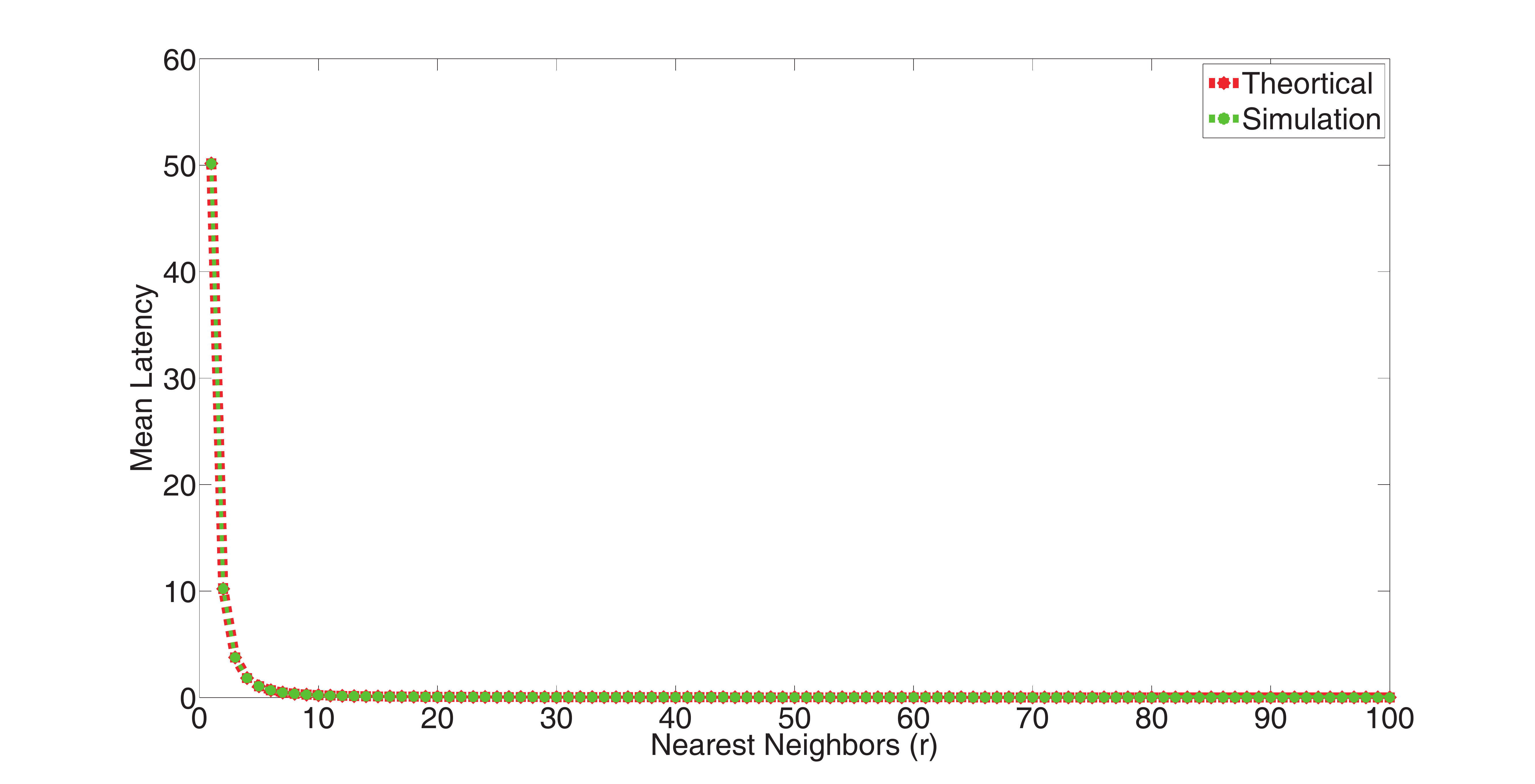}
\caption{Mean latency versus Nearest neighbors for $r$-nearest neighbor cycle}
\end{figure}

\begin{figure}[!t]
\centering
\includegraphics[width=2.5 in]{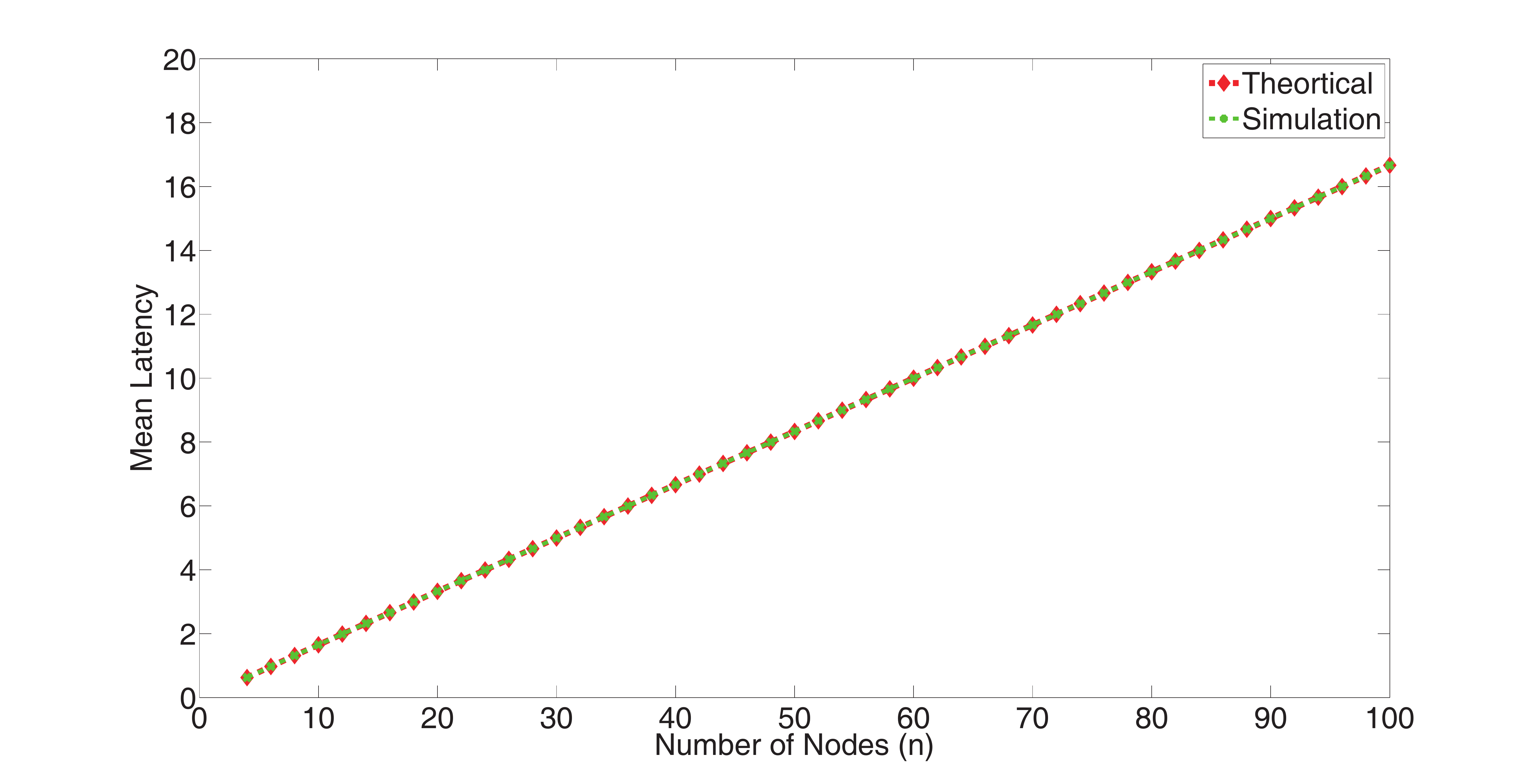}
\label{fig:5}
\caption{Mean latency versus number of nodes for $r$-nearest neighbor cycle}
\label{fig:7}
\end{figure}

\begin{figure}[!t]
\centering
\includegraphics[width=2.5 in]{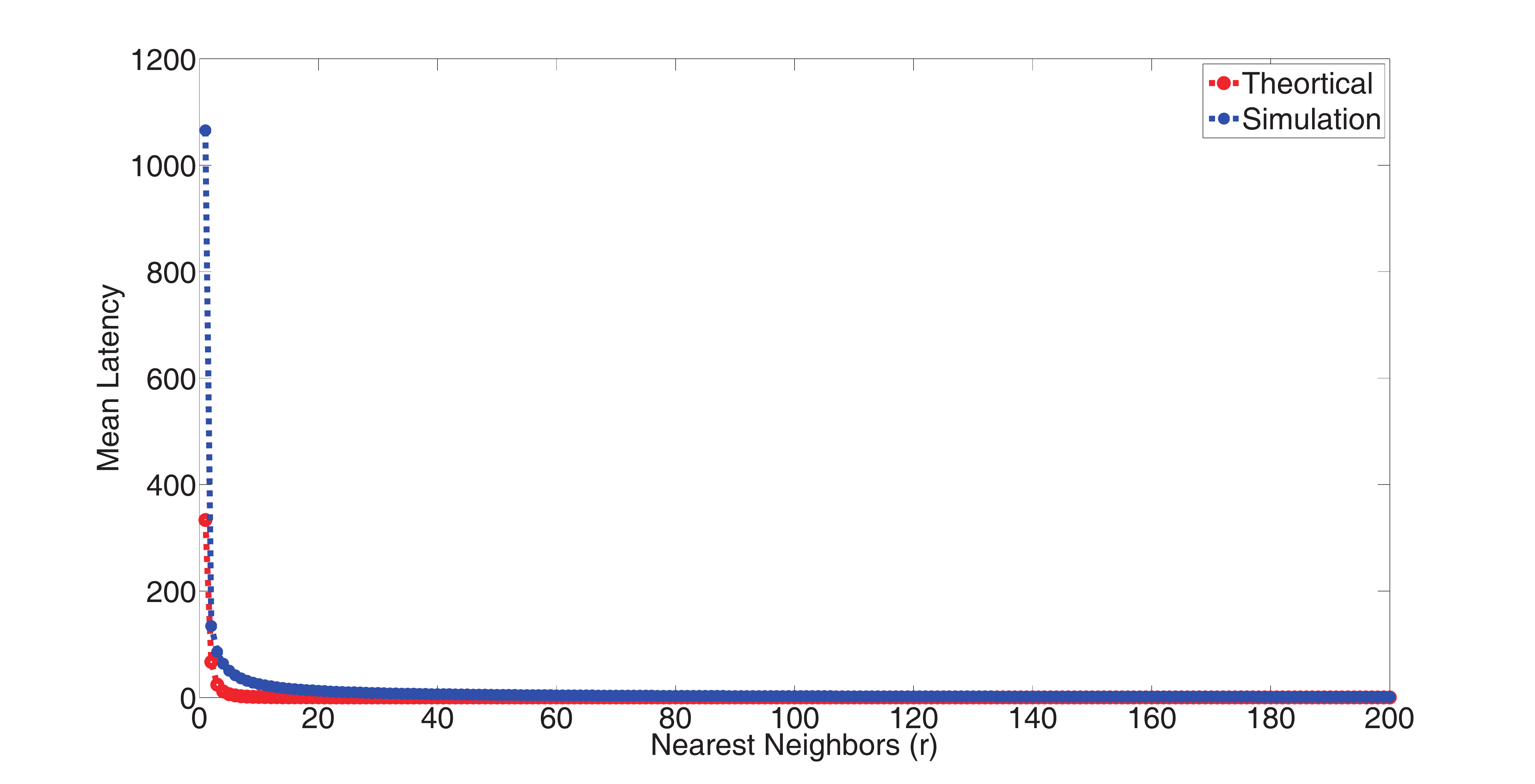}
\caption{Mean latency versus Nearest neighbors for $r$-nearest neighbor torus}
\label{fig:9}
\end{figure}

\begin{figure}[!t]
\centering
\includegraphics[width=2.5 in]{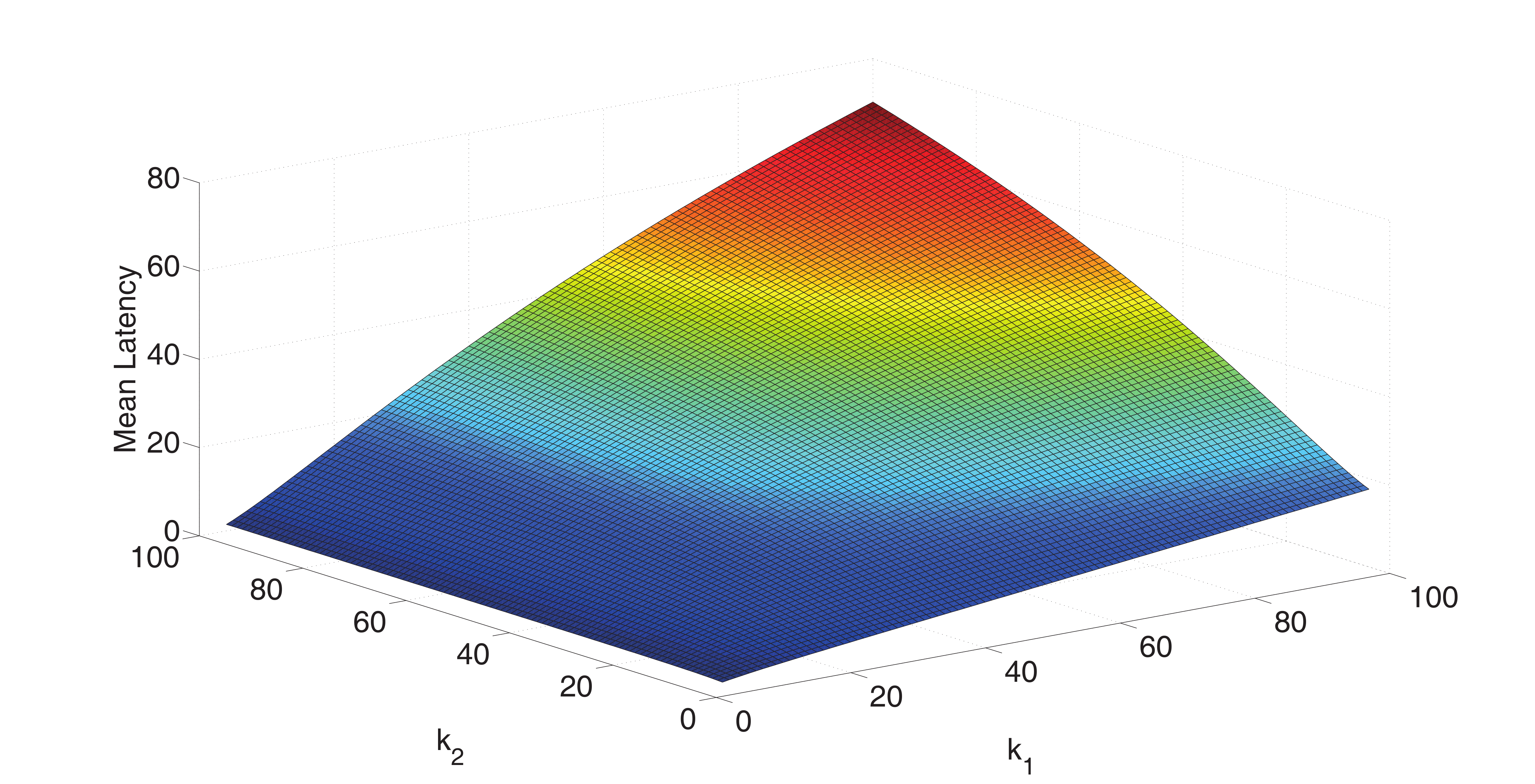}
\caption{Mean latency versus $k_{1}$ and $k_{2}$ for $r$-nearest neighbor torus}
\label{fig:11}
\end{figure}

\begin{figure}[!t]
\centering
\includegraphics[width=2.5 in]{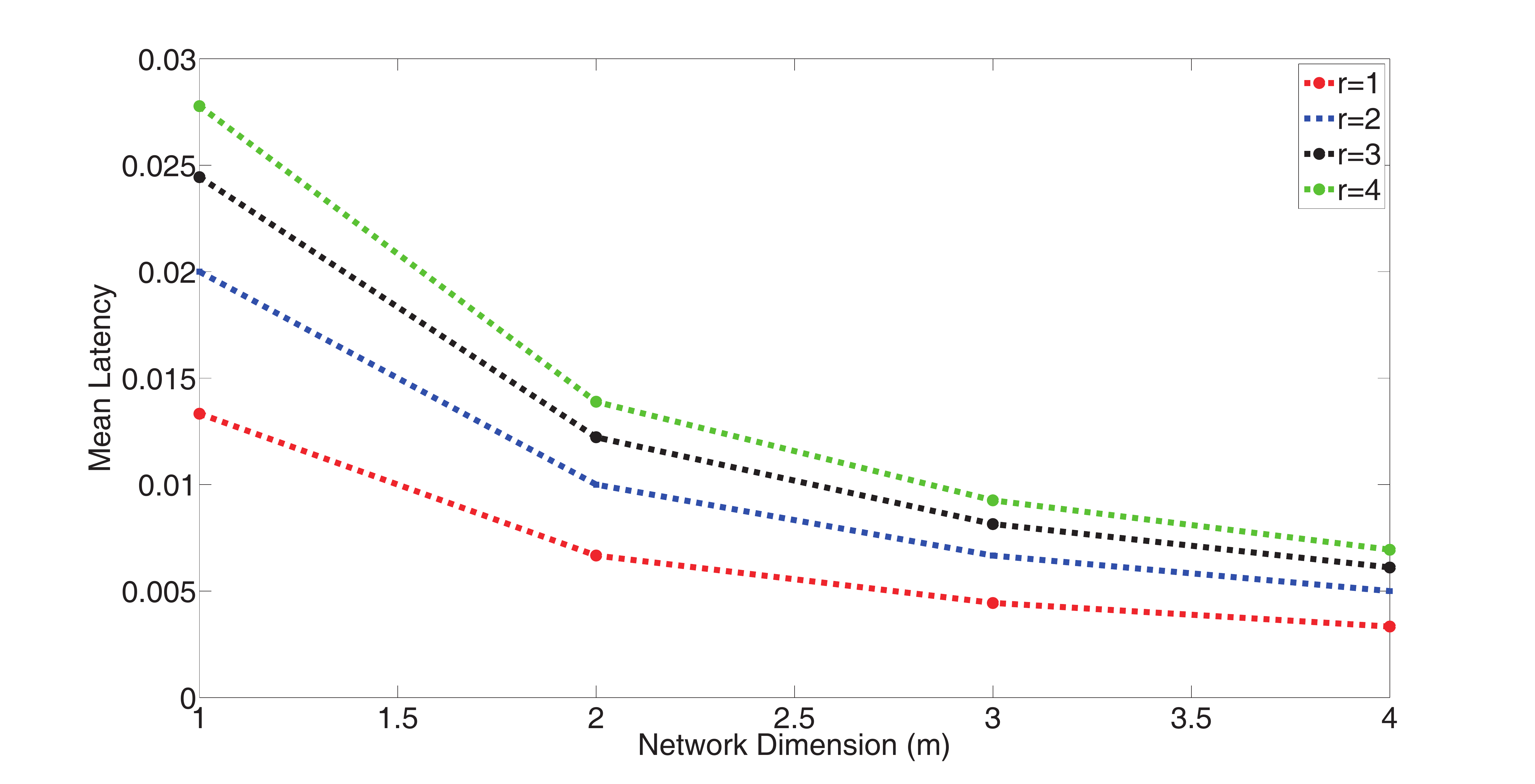}
\caption{Mean latency versus Network Dimension for $r$-nearest neighbor torus}
\label{fig:13}
\end{figure}
\begin{figure}[!t]
\centering
\includegraphics[width=2.5in]{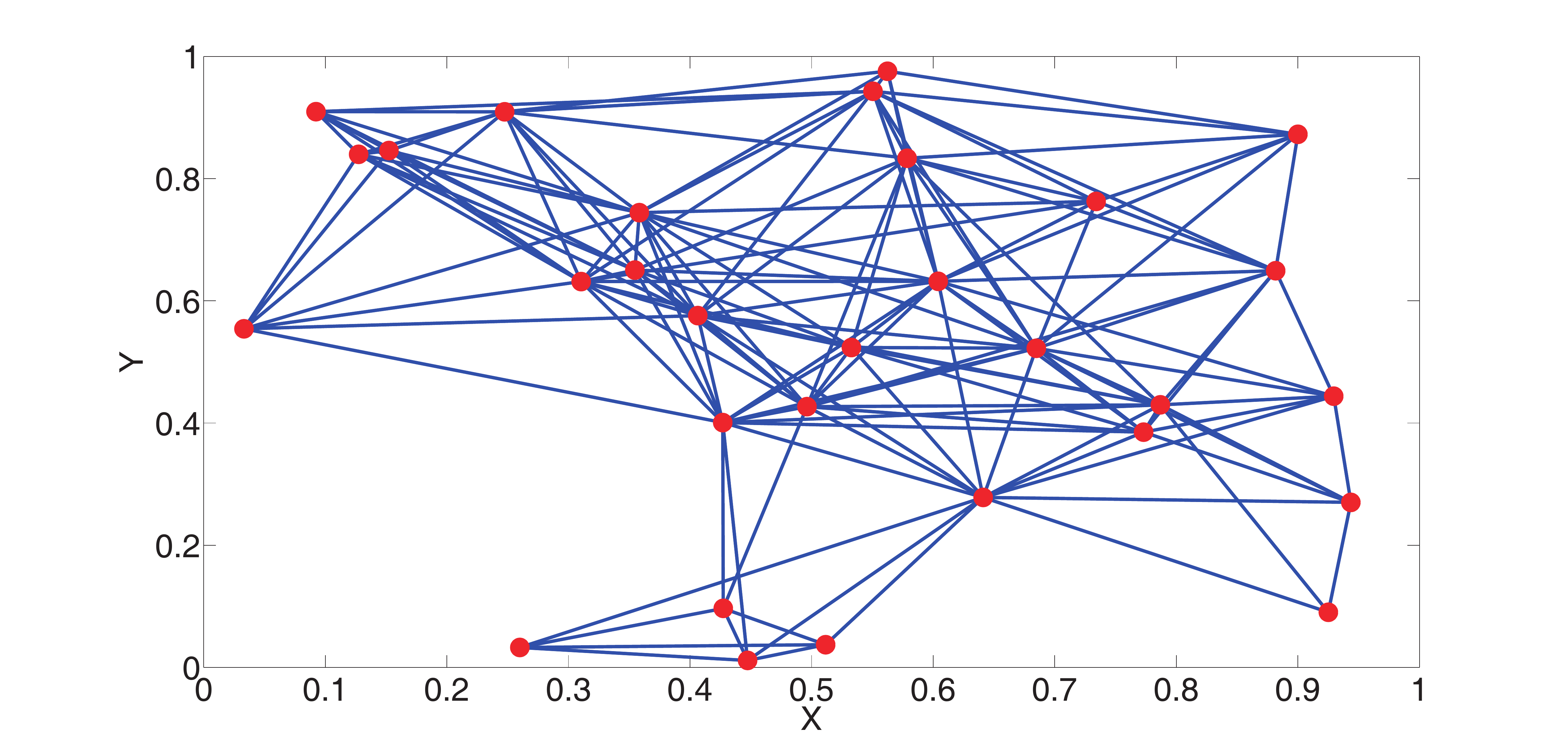}
\caption{Wireless Network Topology}
\label{llklk}
\end{figure}
\begin{figure}[!t]
\centering
\includegraphics[width=2.5in]{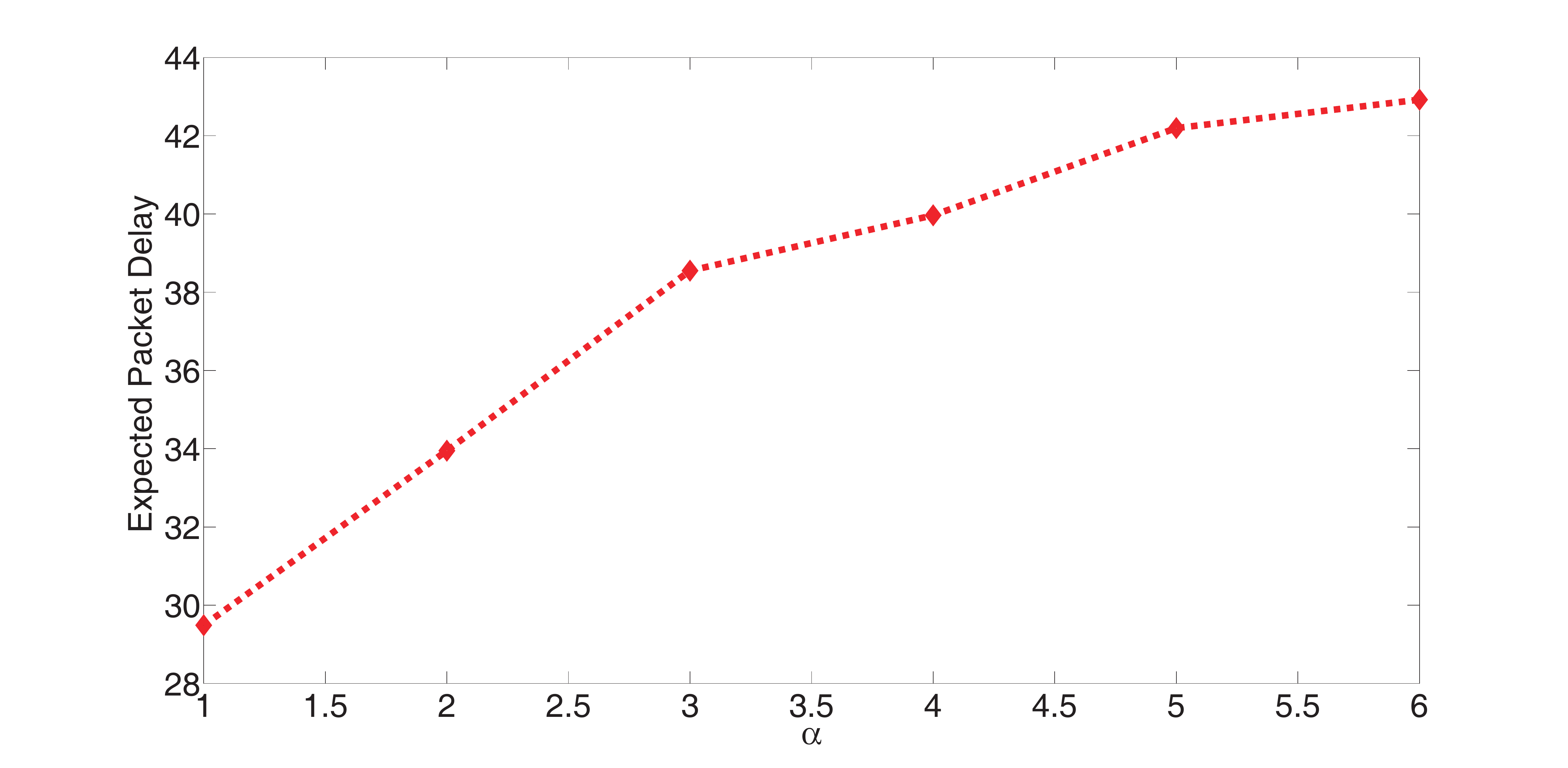}
\caption{Expected Packet Delay versus path loss exponent $\alpha$}
\label{llklk}
\end{figure}
%\begin{figure}[!t]
%\centering
%\includegraphics[width=2.5in]{figure6}
%\caption{Expected Packet Delay versus node density for different topologies}
%\label{llklk}
%\end{figure}
\begin{figure}[!t]
\centering
\includegraphics[width=2.5in]{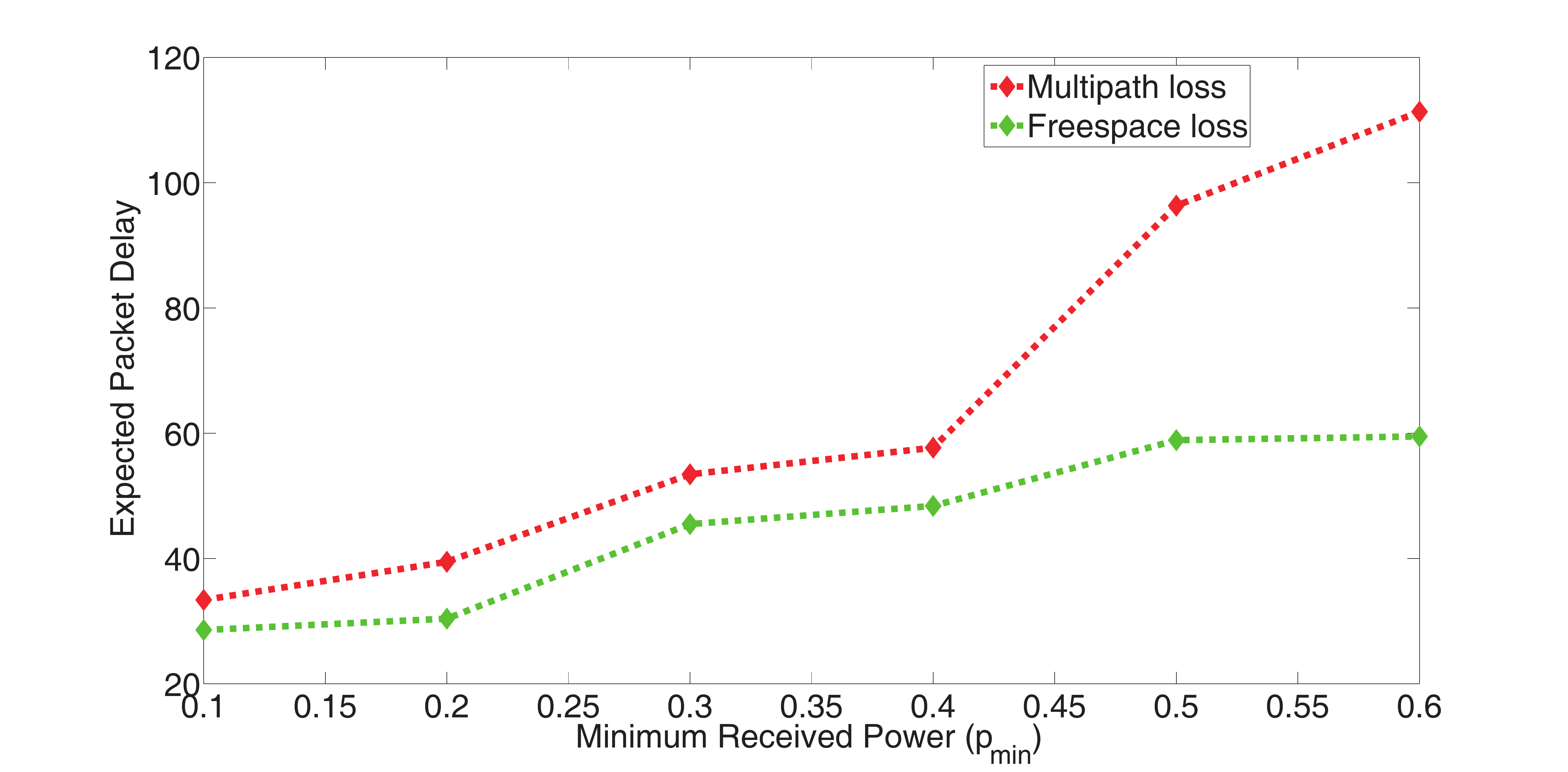}
\caption{Expected Packet Delay versus $P_{min}$ for random topology}
\label{llklk}
\end{figure}

\begin{figure}[!t]
\centering
\includegraphics[width=2.5in]{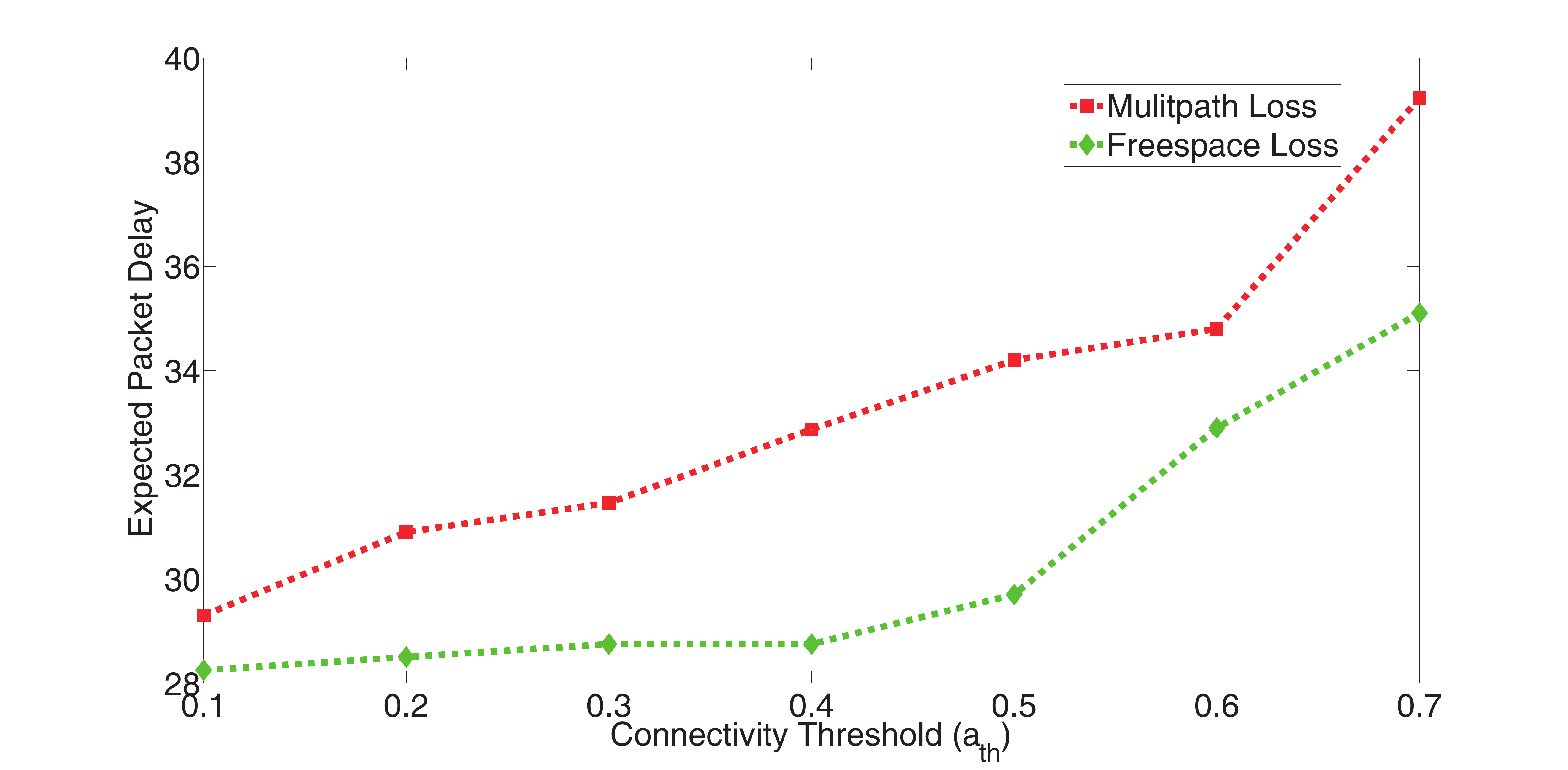}
\caption{Expected Packet Delay versus Connectivity Threshold for random topology}
\label{llklk}
\end{figure}

\section{Conclusions}
In this work, analytical expressions for Mean latency or average travel time has been derived for $r$-nearest neighbor cycle, $r$-nearest neighbor torus and $m$ dimensional $r$ nearest neighbor torus networks. We have given the theoretical bounds for mean latency in terms of number of nodes and nearest neighbors. Further, we also studied the latency for arbitrary wireless networks in flat fading environments.

% that's all folks
\end{document}